\documentclass[a4paper,twoside,prd,preprintnumbers,longbibliography,twocolumn,nofootinbib]{revtex4-1}
\usepackage{amssymb,amsmath,bm,natbib}
\usepackage{cancel}
\usepackage[normalem]{ulem}
\usepackage{xcolor}
\usepackage{slashed}
\usepackage{graphics}
\usepackage{graphicx}
\usepackage[utf8]{inputenc}
\usepackage{tabularx}
\usepackage[caption=false]{subfig}
\usepackage{hyperref}
\usepackage{url}
\usepackage{dsfont}
\usepackage{float} 

\newcommand{\eq}[1]{Eq.~\eqref{#1}}

\renewcommand{\i}{\ensuremath{\mathrm{i}}}

\newcommand{\C}[1]{{\cal C}_{#1}}

\begin{document}
\preprint{PSI-PR-22-02,   ZU-TH 02/22}
\title{Unified Explanation of the Anomalies in Semi-Leptonic \texorpdfstring{$B$}{B} decays and the \texorpdfstring{$W$}{W} Mass}

\author{Marcel Alguer\'{o}}
\email{malguero@ifae.es}
\affiliation{Grup de Fisica   Te\`orica (Departament de Fisica), Universitat Aut\`onoma de Barcelona, E-08193 Bellaterra (Barcelona)}
\affiliation{Institut de Fisica d'Altes Energies (IFAE),
	The Barcelona Institute of Science and Technology, Campus UAB, E-08193 Bellaterra (Barcelona)}

\author{Andreas Crivellin}
\email{andreas.crivellin@cern.ch}
\affiliation{Physik-Institut, Universit\"at Z\"urich, Winterthurerstrasse 190, CH--8057 Z\"urich, Switzerland}
\affiliation{Paul Scherrer Institut, CH--5232 Villigen PSI, Switzerland}

\author{Claudio Andrea Manzari}
\email{claudioandrea.manzari@physik.uzh.ch}
\affiliation{Physik-Institut, Universit\"at Z\"urich, Winterthurerstrasse 190, CH--8057 Z\"urich, Switzerland}
\affiliation{Paul Scherrer Institut, CH--5232 Villigen PSI, Switzerland}

\author{Joaquim Matias}
\email{matias@ifae.es}
\affiliation{Grup de Fisica   Te\`orica (Departament de Fisica), Universitat Aut\`onoma de Barcelona, E-08193 Bellaterra (Barcelona)}
\affiliation{Institut de Fisica d'Altes Energies (IFAE),
	The Barcelona Institute of Science and Technology, Campus UAB, E-08193 Bellaterra (Barcelona)}

\begin{abstract}
The discrepancies between the measurements of rare (semi-)leptonic $B$ decays and the corresponding Standard Model predictions point convincingly towards the existence of new physics for which a heavy neutral gauge boson ($Z^\prime$) is a prime candidate. However, the effect of the mixing of the $Z^\prime$ with the SM $Z$, even though it cannot be avoided by any symmetry, is usually assumed to be small and thus neglected in phenomenological analyses. In this letter we point out that a mixing of the  naturally expected size leads to lepton flavour universal contributions, providing a very good fit to $B$ data. Furthermore, the global electroweak fit is affected by $Z-Z^\prime$ mixing where the tension in the $W$ mass, recently confirmed and strengthened by the CDF measurement, prefers a non-zero value of it. We find that a $Z^\prime$ boson with a mass between $\approx 1-5\rm {TeV}$ can provide a unified explanations of the $B$ anomalies and the $W$ mass. This strongly suggests that the breaking of the new gauge symmetry giving raise to the $Z^\prime$ boson is linked to electroweak symmetry breaking with intriguing consequences for model building.
\end{abstract}
\maketitle

\newpage
\section{Introduction}

Even though the LHC has not discovered any particles beyond the ones of the Standard Model (SM) yet, in the last years intriguing hints for the violation of lepton flavour universality (LFU) have been accumulated (see e.g.~Refs.~\cite{Crivellin:2021sff,Fischer:2021sqw,Crivellin:2022qcj} for recent reviews). Among them, the updated measurement of the ratios of semi-leptonic rare $B$ meson decay $R_{K^+}={\cal B}(B^+\to K^+\mu^+\mu^-)/{\cal B}(B^+\to K^+ e^+e^-)$~\cite{Hiller:2003js} by LHCb~\cite{LHCb:2021trn} is particularly prominent since it provides first evidence for LFU violation (LFUV) in a single observable. Furthermore, when combining all tests of LFUV (like $R_{K^+}$)~\cite{LHCb:2017avl,Capdevila:2016ivx,Belle:2016fev,LHCb:2021lvy} with $B$ decays involving muon pairs (most prominently $P_5^\prime$~\cite{Descotes-Genon:2012isb,LHCb:2020lmf} and $B_s\to\phi\mu^+\mu^-$~\cite{LHCb:2021xxq,LHCb:2021zwz}), one finds a preference for new physics (NP) hypotheses of more than $7\,\sigma$~\cite{Alguero:2021anc} 
compared to the SM~\footnote{Very close results and pulls were found in Ref.~\cite{Hurth:2021nsi} using also a complete set of observables but a different treatment of hadronic uncertainties. See also Ref.~\cite{Altmannshofer:2021qrr} for an analysis using a smaller subset of the available data as well as Refs.~\cite{Kowalska:2019ley,Blake:2019guk,Geng:2021nhg,Ciuchini:2019usw} and Ref.~\cite{London:2021lfn} for a detailed comparison.}. Note that such a high significance is only possible since all measurements are compatible with each other, i.e.~they form a coherent picture.

Simple patterns where NP couples solely to muons can in fact explain the discrepancies between the SM and experiment in rare semi-leptonic $B$ decays very well. However, it turns out that structures with additional LFU contributions can describe data even better~\cite{Alguero:2018nvb,Alguero:2022wkd}. This means that allowing simultaneously for presence of LFUV and LFU NP effects, one can further improve the goodness of the global fits. Indeed, some of these hypotheses exhibit the highest significance among all studied scenarios~\cite{Capdevila:2017bsm,Alguero:2019ptt,Alguero:2021anc,Altmannshofer:2021qrr}.\footnote{In fact, several models giving raise to combined LFU and LFUV contributions, including 2HDMs~\cite{Crivellin:2019dun}, leptoquarks~\cite{Crivellin:2018yvo,Crivellin:2019dwb}, $SU(2)_L$ triplets vector bosons~\cite{Capdevila:2020rrl} and models with vector-like quarks~\cite{Bobeth:2016llm,Crivellin:2020oup}, have been proposed in the literature.}

In this letter, we point out that, extending the SM by a new heavy neutral gauge boson ($Z^\prime$), one has, in addition to the usually considered direct LFUV effect in $b\to s\ell^+\ell^-$~\cite{Buras:2013qja,Gauld:2013qba,Gauld:2013qja,Altmannshofer:2014cfa,Crivellin:2015mga,Crivellin:2015lwa,Niehoff:2015bfa,Sierra:2015fma,Carmona:2015ena,Falkowski:2015zwa,Celis:2015eqs,Celis:2015ara,Crivellin:2015era,Boucenna:2016wpr,Altmannshofer:2016oaq,Boucenna:2016qad,Crivellin:2016ejn,GarciaGarcia:2016nvr,Faisel:2017glo,King:2017anf,Chiang:2017hlj,DiChiara:2017cjq,Ko:2017lzd,Sannino:2017utc,Carmona:2017fsn,Raby:2017igl,Falkowski:2018dsl,Benavides:2018rgh,Maji:2018gvz,Singirala:2018mio,Guadagnoli:2018ojc,Allanach:2018lvl,Kohda:2018xbc,King:2018fcg,Duan:2018akc,Rocha-Moran:2018jzu,Dwivedi:2019uqd,Foldenauer:2019vgn,Ko:2019tts,Allanach:2019iiy,Kawamura:2019rth,Altmannshofer:2019xda,Calibbi:2019lvs,Aebischer:2019blw,Kawamura:2019hxp,Crivellin:2020oup,Allanach:2020kss,Greljo:2021xmg,Davighi:2021oel,Bause:2021prv,Allanach:2021kzj,Navarro:2021sfb,Ko:2021lpx,Allanach:2021gmj}, also a LFU effect, which is generated via $Z-Z^\prime$ mixing. In fact, because both bosons have the same quantum numbers, this mixing cannot be avoided by any symmetry. Furthermore, in the case that electroweak (EW) symmetry breaking and the breaking of the symmetry giving rise to the $Z^\prime$ mass are connected, one even expects a mixing of the order of $m_Z^2/m_{Z^\prime}^2$. Importantly, $Z-Z^\prime$ mixing has also an impact on the global EW fit, in particular on $Z\ell^+\ell^-$ and $Z\nu\nu$ couplings and if the $Z^\prime$ is an $SU(2)_L$ singlet (i.e. not the neutral component of an $SU(2)_L$ multiplet), in addition the prediction of the $W$ mass is altered compared to the SM. The latter is very important since the global EW fit displayed a tension of $1.8\,\sigma$~\cite{deBlas:2021wap} in this observable. This discrepancy was recently confirmed and strengthened by the CDF measurement~\cite{CDF:2022hxs} whose central value is $7\,\sigma$ above the SM prediction~\cite{ParticleDataGroup:2020ssz}. Combining this new measurement with the existing ones from the LHC~\cite{ATLAS:2017rzl,CMS:2011utm,LHCb:2015jyu,LHCb:2021bjt},  one finds $m_W=(80.4133\pm 0.0080)$GeV and $m_W=(80.413\pm 0.015)$GeV, where in the second formula the error has been inflated to reflect the tensions between the different measurements. The SM prediction is given by $m_W^{\rm SM}=(80.3499\pm 0.0056)$GeV, and $m_W^{\rm SM}=(80.3505\pm 0.0077)$GeV for a conservative error estimate~\cite{deBlas:2022hdk}. This corresponds to a $6.5\,\sigma$ and $3.7\,\sigma$ tension for the standard and the conservative scenario, respectively.\footnote{Note that this article was submitted before the $W$ mass measurement of CDF was released such that this result can be considered as a confirmation of the prediction of the original version of the manuscript.}

Therefore, in $Z^\prime$ models an interesting interplay between $b\to s\ell^+\ell^-$ processes and the global EW fit arises if the $Z-Z^\prime$ mixing angle is non-zero~\cite{Allanach:2021kzj}. While this mixing has usually been assumed to be negligibly small~\footnote{Note that the effect of $Z-Z^\prime$ mixing in the $W$ mass in the context of $b\to s\ell^+\ell^-$ was already pointed out in Ref.~\cite{Allanach:2021kzj}. However, the impact on, and the correlations with, the $b\to s\ell^+\ell^-$ fit were not shown. 
}, the goal of this letter is to assess the size and impact of $Z-Z^{\prime}$ mixing via a combined analysis of flavour and EW data, providing a unified explanation of both anomalies.

\section{Setup}
\label{Setup}

We extend the SM by adding a heavy neutral $SU(2)_L$ singlet gauge boson. Following the notation of Ref.~\cite{delAguila:2010mx,deBlas:2012qp} the kinetic term and the mass term of this new boson, before EW symmetry breaking, are
\begin{align}
\begin{aligned}
	\mathcal{L}_{Z'_0}=&-\frac{1}{4}Z'_{0,\mu\nu}Z_0^{\prime\mu\nu}+\frac{\mu_Z^{\prime2}}{2}Z^\prime_{0\mu} Z^{\prime\mu}_0\,\\
	&+g_{Z'}Z'_{0\mu} Z'^\mu_0\phi^\dagger \phi 
	-i g_{Z'}^\phi Z'^\mu_0\phi^\dagger \overleftrightarrow{D}_\mu \phi\,,\label{LZp}
	\end{aligned}
\end{align}
where $Z'_{0,\mu\nu}\equiv \partial_\mu Z'_{0\nu}-\partial_\nu Z'_{0\mu}$ is the field strength tensor, $\overset{\leftrightarrow}{D}_{\mu}\; =\overset{\rightarrow}{D}_{\mu}- (\overset{\leftarrow}{D}_{\mu})^\dagger$, $\phi$ is the SM Higgs $SU(2)_L$ doublet and we use
$D_\mu=\partial_\mu + ig_2W^a_\mu T^a+ig_1YB_\mu $
as the definition of the SM part of the covariant derivative and $g_{Z'}^\phi$ is real by hermicity. The physical $Z$ and $Z^\prime$ masses are obtained from diagonalizing the mass matrix
\begin{equation}
	\mathcal{M}^2=\begin{pmatrix}
		m_{Z_0}^2 & -\frac{y}{c_W}\\
		-\frac{y}{c_W}& m_{Z'_0}^2
	\end{pmatrix}\,, \qquad    y\equiv \frac{v^2}{2}\,g_2 \,g_{Z'}^\phi\,,
\end{equation}
in the $Z_0,\,Z_0'$ basis, where $Z_0$ coincides with the the SM $Z$ for $g_{Z'}^\phi=0$ with $m_{Z_0}^2=\frac{v^2}{4}\left(g_1^2+g_2^2\right)$, $\frac{v}{\sqrt{2}}\approx 174\,$GeV and $c_W$ is the cosine of the Weinberg angle. At leading order in $v/m_{Z_0^\prime}$ we have
\begin{align}
	m_Z^2\simeq & \;m_{Z_0}^2-\frac{y^2}{c_W^2 m_{Z_0 '}^2}\equiv m_{Z_0}^2\left(1+\delta m_Z^2\right)\,.\label{MZcorr}
\end{align}
Note that the corrections to the mass of the $Z$ with respect to the SM value $m_{Z_0}$ can only be negative. The mass eigenstates $Z^{(\prime)}$ can then be expressed as
\begin{equation}
	\left( {\begin{array}{*{20}{c}}
			Z \\	{Z^\prime}
	\end{array}} \right) = \left( {\begin{array}{*{20}{c}}
	    	{Z_{0}^{\prime} \sin \xi + Z_{0} \cos \xi }\\
			{Z_{0}^{\prime} \cos \xi \; - Z_{0} \sin \xi}
	\end{array}} \right)\,,
	\label{eq:mixing}
\end{equation}
where $
	\sin \xi\simeq \frac{y}{c_W m_{Z_0'}^2}$
describes the $Z-Z'$ mixing. 

The interactions with the SM fields are given by
\begin{align}
	\begin{split}
	\mathcal{L}_{Z'_0}^{\rm fermions}=	&{\bar u}_j{\gamma _\mu }(g_{ji}^{uL}{P_L} + g_{ji}^{uR}{P_R}){u_i} \,Z^{\prime\mu}_0 \\
	& +{\bar d}_j {\gamma _\mu }(g_{ji}^{dL} {P_L}+g_{ji}^{dR} {P_R}){d_i} \, Z^{\prime\mu}_0 \\
	& + g_{ji}^{\ell L} ({\bar \nu}_j {\gamma _\mu }{P_L}{\nu_i}) \, Z^{\prime\mu}_0 \\
	& + \bar {\ell}_j {\gamma _\mu }(g_{ji}^{\ell L} {P_L}+ g_{ji}^{\ell R} {P_R}){\ell_i} \, Z^{\prime\mu}_0 \,,
	\end{split}
	 \label{Lint}
\end{align}
where, in the down basis, $g_{ji}^{uL}=V_{jk} g_{kk'}^{dL}V_{ik'}^*$. Note that the couplings to  left-handed charged leptons and neutrinos (up and down quarks) are the same (up to a CKM rotation), due to $SU(2)_L$ invariance and that only the relative phase between $\sin \xi$ and $g^{L,R}_{ij}$ is physical, such that one can assume $\sin\xi$ to be positive without loss of generality. In the following, we will assume flavour diagonal coupling to leptons and in the quark sector disregard all couplings except left-handed $b-s$ couplings.

\section{Observables}
\label{Observables}

\subsection{\texorpdfstring{$b\to s\ell^+\ell^-$}{b->sll}}

\begin{table}[t]
	\begin{tabular}{cccc}
		\toprule 
		& Best-fit point & $1\,\sigma$ CI & $2\,\sigma$ CI \\
		\hline
		$\C{9\mu}^{\rm V}$ & $-0.96$ & $[-1.11,-0.80]$ & $[-1.25,-0.64]$\\
		$\C{10}^{\rm U}=-k \C{9}^{\rm U}$ & $+0.30$ & $[+0.15,+0.45]$ & $[+0.00,+0.61]$\\
		\hline
	\end{tabular} 
	\caption{ $1\,\sigma$ and $2\,\sigma$ confidence intervals for the NP scenario in \eq{bsll_sc} with a Pull$_{\rm SM}$ of $6.9\,\sigma$ and p-value=28.3\%.}
	\label{ZpSc1}
\end{table} 
\begin{figure}
	\centering 
	\includegraphics[width=0.38\textwidth]{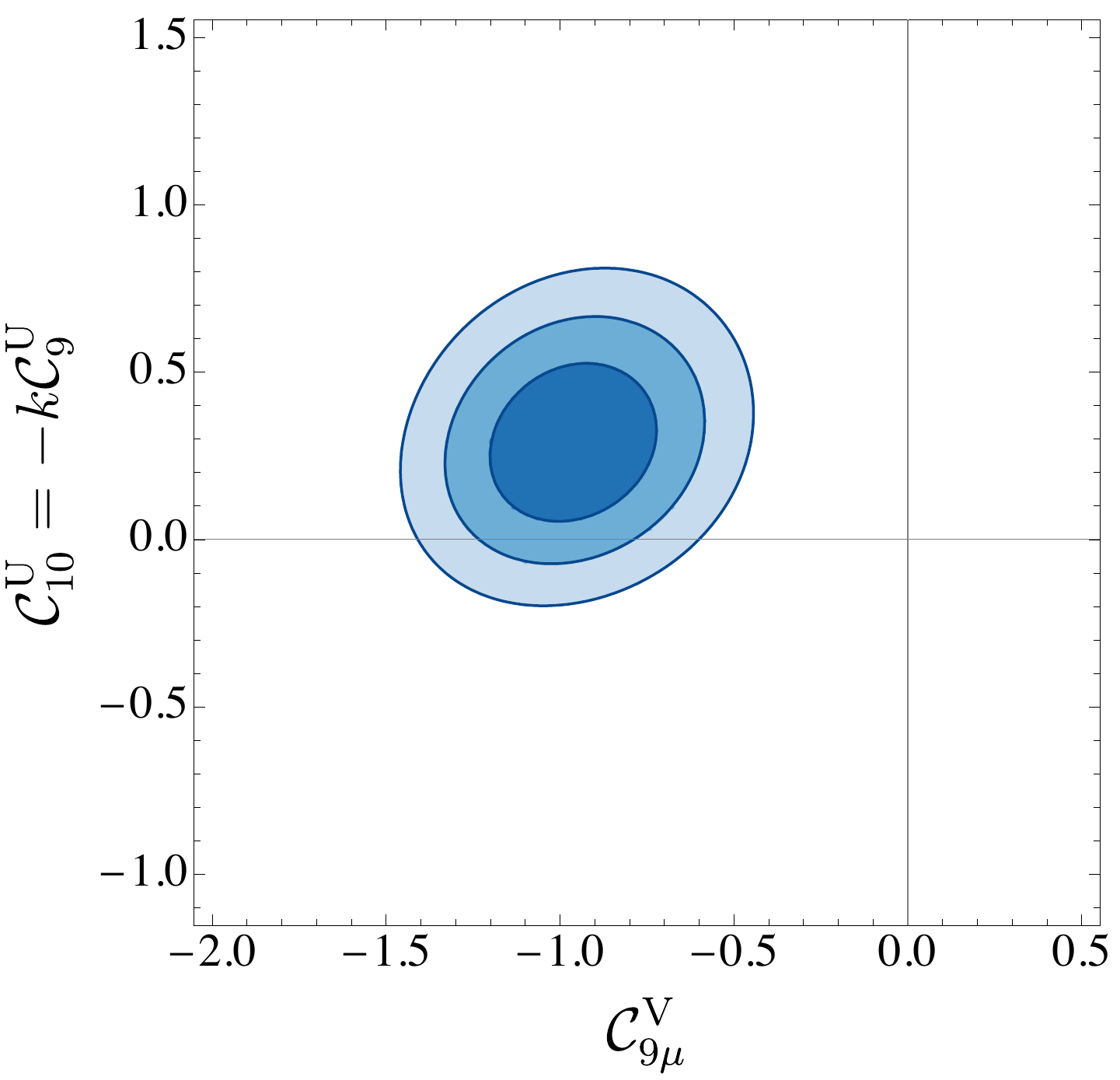}
	\caption{Preferred 1$\,\sigma$, 2$\,\sigma$ and 3$\,\sigma$ regions in the $(\C{9\mu}^{\rm V},\C{10}^{\rm U}=-k \C{9}^{\rm U})$ plane for the scenario {discussed in the paper}, including all available $b\to s\ell^{+}\ell^{-}$ data and using the most updated version of ACDMN code~\cite{Alguero:2021anc}. Note that the SM case corresponds to the (0,0) point.}%
	\label{fig:bsll_fit}%
\end{figure}

In $Z^\prime$ models without $Z-Z^\prime$ mixing, the simple one dimensional scenario with the best fit to data is obtained from a left-handed $b-s$ coupling and a vectorial muon coupling, i.e. the ${\cal C}_{9\mu}^{\rm V}$ scenario~\cite{Alguero:2021anc}. Allowing in addition for $Z-Z^\prime$ mixing we have
\begin{equation} \label{eq6}
\begin{aligned}
{\cal C}_{9\mu}^{\rm V} &=  - \frac{\pi ^2}{e^2}\frac{4\sqrt{2} g_{23}^{dL} g_{22}^{\ell V} }{ G_F m_{Z^{\prime}}^2 V_{tb} V_{ts}^*}\,,\\
		{\cal C}_{10}^{{\rm U}} &=-k {\cal C}_{9}^{\rm U}=
		\dfrac{\sqrt{2}\pi ^2}{e^2}\frac{g_2\, g_{23}^{dL} \sin\xi }{ c_W G_Fm_Z^2 V_{tb}V_{ts}^*}\,,
\end{aligned}
\end{equation}
using the effective Hamiltonian of Ref.~\cite{Grinstein:1987vj,Buchalla:1995vs} where $g_{22}^{\ell V} =( {g_{22}^{\ell L} + g_{22}^{\ell R}})/{2}$. This corresponds to the scenario
\begin{equation}
\{{\cal C}_{9\mu}^{\rm V}\,,\, {\cal C}_{10}^{\rm U}= -k{\cal C}_{9}^{\rm U}\}\,,
\label{bsll_sc}
\end{equation}
with $k=1/(1-4 s_w^2)$ (see the appendix for the definitions of the operators). The superscript ${\rm V}$ (${\rm U}$) in the Wilson coefficient stands for a LFUV (LFU) contribution.

We perform the most recent fit~\cite{Alguero:2021anc} to the scenario in Eq.~\eqref{bsll_sc}, including 254 observables and the latest measurements by LHCb of LFUV observables, namely, $R_{K^0_S}$~\cite{LHCb:2021lvy} and $R_{K^{*+}}$~\cite{LHCb:2021lvy} as well as the new branching ratio and angular distribution of $B_s \to \phi\mu^+\mu^-$~\cite{LHCb:2021xxq,LHCb:2021zwz}. We obtain the best fit point and confidence level regions in Table~\ref{ZpSc1}. The results of the global fit in our scenario are shown in Fig.~\ref{fig:bsll_fit}.

\subsection{\texorpdfstring{$B_s-\bar B_s$ Mixing}{Bs-Bs}}

The most important constraint on $Z^\prime-b-s$ couplings, i.e. $g_{23}^{dL}$, comes from $B_s-\bar B_s$ mixing where the contribution to the Hamiltonian ${\cal H}_{\rm eff}={\cal C}_1 {\cal O}_1$, with 
$${\cal O}_1=\big(\bar b \gamma^\mu P_L s\big)\times \big(\bar b \gamma_\mu P_L s\big),$$ is given by
\begin{equation}{{\cal C}_1} = \frac{1}{2}\bigg(\frac{g_{23}^{dL}}{m_{Z^{\prime}}}\bigg)^2\left( 1 + \frac{\alpha _s}{4\pi}\frac{11}{3} \right)\,,\end{equation}
including the NLO matching corrections of Ref.~\cite{Buras:2012fs}. Note that the effect of the mixing induced $Z-b-s$ couplings can be neglected as it corresponds to a dimension 8 contribution. 
Employing  the 2-loop renormalization group evolution~\cite{Ciuchini:1997bw,Buras:2000if}, this leads to an effect, normalized to the SM one, of

$$\left(\frac{g^{dL}_{23}}{0.52}\right)^2 \left(\frac{10 {\rm TeV}}{m_{Z^\prime}} \right)^2=0.110\pm0.090$$
using the bag factor of Ref.~\cite{Aoki:2019cca} and the global fit to NP in $\Delta F=2$ observables of Ref.~\cite{Bona:2007vi}. 
\subsection{LFUV in tau decays}

Assuming lepton flavour conservation, $Z^\prime-W$ boxes contribute to $\tau \to \mu \,\nu_\tau \overline{\nu}_\mu$ as~\cite{Altmannshofer:2014cfa}:
\begin{align}
\begin{split}
\frac{\mathcal{A}(\tau \to \mu \,\nu_\tau \overline{\nu}_\mu)}{\mathcal{A}(\tau \to \mu \,\nu_\tau \overline{\nu}_\mu)_{SM}}
=&1-\frac{3}{8\pi^2}\,g^{\ell L}_{22}\,g^{\ell L}_{33} \frac{\ln \left(\frac{m_W^2}{m_{Z^{\prime}}^2}\right)}{1-\frac{m_{Z^{\prime}}^2}{m_W^2}}\,,
\end{split}
\label{taumununu}
\end{align}
and analogously for $\tau \to e \,\nu_\tau \overline{\nu}_e$ and $\mu \to e \,\nu_\mu \overline{\nu}_e$. Note that at vanishing momentum transfer the $Z^\prime$ induced correction to the $W$-$\ell$-$\nu$ vertex vanishes as $SU(2)_L$ gauge invariance is not broken. This we compared to the experimental results~\cite{Amhis:2019ckw} (see Ref.~\cite{Bryman:2021teu} for an overview on LFUV):
\begin{align}
\label{eq:LFUratios}
	\begin{split}
\left.\frac{\mathcal{A}\left[\tau  \to \mu \nu \bar \nu \right]}{\mathcal{A}\left[\mu  \to e\nu \bar \nu \right]}\right|_{\rm EXP} &= 1.0029 \pm 0.0014\,,\\
\left.\frac{\mathcal{A}\left[\tau  \to \mu \nu \bar \nu \right]}{\mathcal{A}\left[\tau  \to e\nu \bar \nu \right]}\right|_{\rm EXP} &= 1.0018 \pm 0.0014\,,\\
\left.\frac{\mathcal{A}\left[\tau  \to e\nu \bar \nu \right]}{\mathcal{A}\left[\mu  \to e\nu \bar \nu \right]}\right|_{\rm EXP} &= 1.0010 \pm 0.0014\,,
\end{split}
\end{align}
with the correlation matrix given in Ref.~\cite{Amhis:2019ckw}.\footnote{Here we neglected semi-leptonic tau decays as well as other probes of LFUV in the charged current which are not affected in the absence of quark coupling (see Ref.~\cite{Bryman:2021teu} for a  recent review). }

\subsection{Electroweak fit}

The EW sector of the SM has been tested with a very high precision at LEP~\cite{Schael:2013ita,ALEPH:2005ab} but also at the Tevatron~\cite{CDF:2013dpa} and the LHC~\cite{ATLAS:2017rzl,CMS:2011utm,LHCb:2015jyu}. Since it can be parametrized by only three Lagrangian parameters, we choose as usual the set with the smallest experimental error consisting of the Fermi constant ($G_F=1.1663787(6)\times10^{-5}\,{\rm GeV}^{-2}$~\cite{Zyla:2020zbs}), the mass of the $Z$ boson ($m_Z=91.1875(21)$ GeV~\cite{ALEPH:2005ab}) and the fine structure constant $\alpha_{em}=7.2973525664(17)\times10^{-3}$~\cite{Zyla:2020zbs,Mohr:2015ccw,Bouchendira:2010es,Parker:2018vye}.

In our model, the relation between the Lagrangian values and the measurements of $G_F$ and $m_Z$ is shifted with respect to the SM. While the effect in $\mu \to e\nu \bar \nu $ is analogous to the one in $\tau \to \mu\nu \bar \nu $ discussed above we have 
$\frac{m_Z^2}{m_{Z_0}^2} \approx 1-\sin \xi^2\frac{m_{Z^{\prime}_0}^2}{m_{Z_0}^2}$. However, since the $Z$ mass is used as an input, this translates into a shift in the $W$ mass prediction of approximately
\begin{align}
	\frac{m_W^2}{m_{W_0}^2}\approx 1+\sin \xi^2\frac{m_{Z^{\prime}_0}^2}{m_{Z_0}^2}\,.
		\label{eq:EW:Wmass}
\end{align}
Note that this shift is positive definite such that the corresponding tension can be explained. 

This modification of the $W$ mass as well as $Z\ell\ell$ and $Z\nu\nu$~\cite{ALEPH:2005ab} are implemented in HEPfit~\cite{deBlas:2019okz} (including the $Z^\prime$ vertex corrections~\cite{Altmannshofer:2014cfa,Haisch:2011up}). In addition, the Higgs mass ($m_H = 125.16 \pm 0.13$ GeV~\cite{Aaboud:2018wps,CMS:2020xrn}), the top mass ($m_t = 172.80 \pm 0.40$ GeV~\cite{TevatronElectroweakWorkingGroup:2016lid,Aaboud:2018zbu,Sirunyan:2018mlv}), the strong coupling constant ($\alpha_s(m_Z) = 0.1181\pm 0.0011$~\cite{Zyla:2020zbs}) and the hadronic contribution to the running of $\alpha_{\rm em}$ ($\Delta\alpha_{\rm had}=276.1(11) \times 10^{-4}$~\cite{Zyla:2020zbs}) have been used as input parameters, since they enter EW observables indirectly via loop effects. The complete set of observables used are listed in the appendix.

\begin{figure}
	\centering 
	\includegraphics[width=0.43\textwidth]{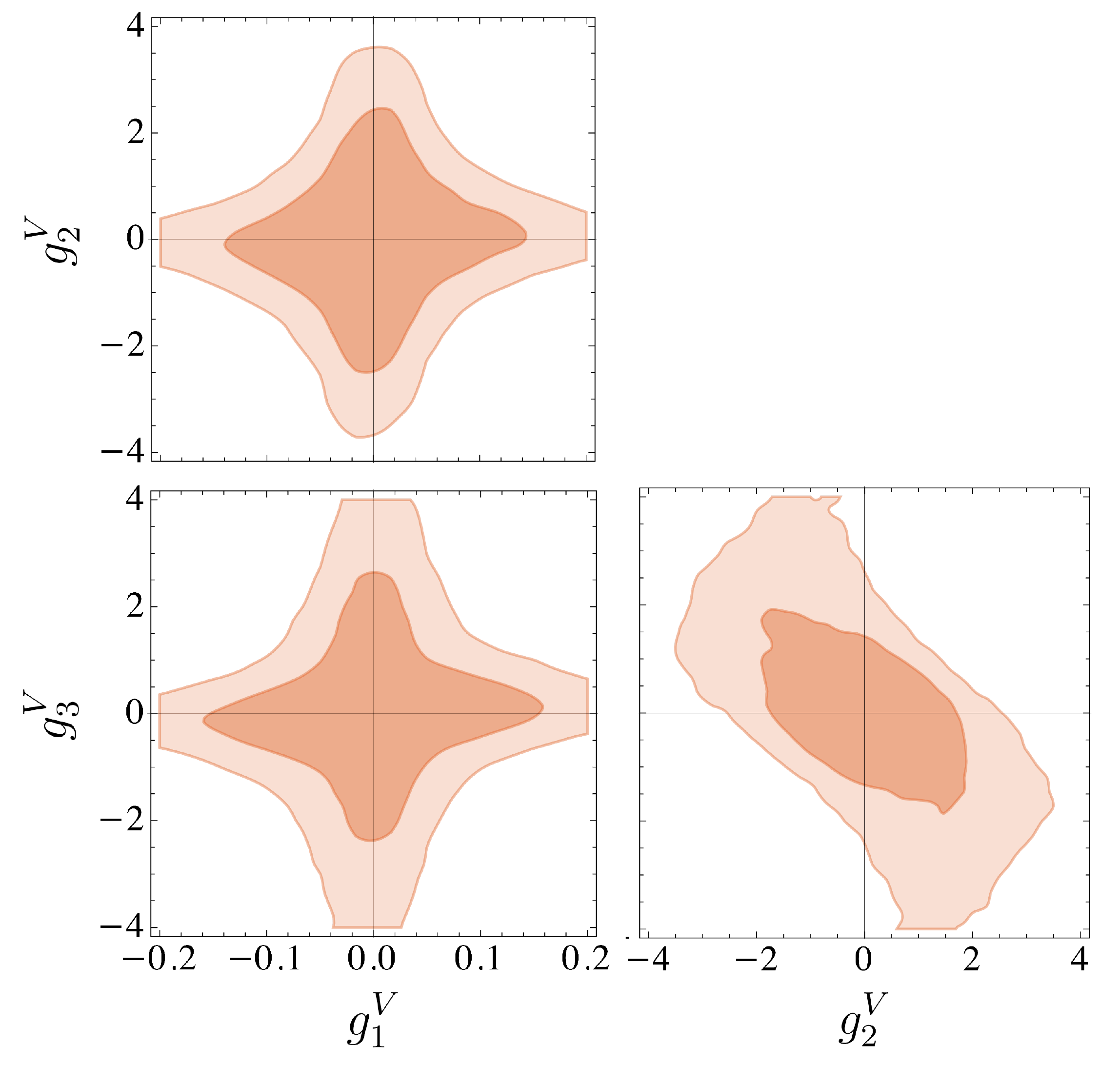}
	\caption{Global fit to EW data, neutrino trident production, LEP bounds on 4-lepton contact interactions and $\tau\to\mu\nu\nu$ data with vectorial flavour diagonal  couplings $g^L_{ii} = g^R_{ii} = g^V_i$. Here we marginalized over the $Z-Z^\prime$ mixing angle $\xi$. The 68\% and 95\% confidence level regions are shown for a $Z^\prime$ mass of $2\,$TeV. Note that a preference for the $L_{\mu}-L_{\tau}$ scenario emerges.}
	\label{fig:EWfit}%
\end{figure}

\subsection{Neutrino Trident Production}

The production of a $\mu^+\mu^-$ pair from the scattering of a muon-neutrino off the Coulomb field of a nucleus, known as neutrino trident production, constitutes a sensitive probe of new neutral current interactions in the lepton sector~\cite{Altmannshofer:2014cfa,Altmannshofer:2014pba}. Generalizing the formula of Ref.~\cite{Altmannshofer:2014pba} we find
\begin{align}
\begin{split}
	&\dfrac{{{\sigma_{\rm SM + NP}}}}{{{\sigma _{\rm SM}}}} = 1\\
	&  + 8 \dfrac{g_{22}^{\ell L}}{g_2^2}\dfrac{m_W^2}{m_{Z'}^2} \dfrac{\left( {1 + 4s_W^2} \right)\left( {g_{22}^{\ell L} + g_{22}^{\ell R}} \right)+ \left( {g_{22}^{\ell L} - g_{22}^{\ell R}} \right)}{\left( {1 + 4s_W^2} \right)^2 + 1}\,.
	\end{split}
\end{align}
This ratio is bounded 
by the weighted average
$\sigma_{\rm exp } / \sigma_{\mathrm{SM}}=0.83 \pm 0.18$
obtained from averaging the CHARM-II~\cite{Geiregat:1990gz}, CCFR~\cite{Mishra:1991bv} and NuTeV results~\cite{Adams:1998yf}.

\subsection{Direct searches}

LEP-II sets stringent bounds on 4-lepton operators from $e^+ e^- \to \ell^+ \ell^-$ (with $\ell = e, \mu, \tau$)~\cite{Schael:2013ita} for specific chiralities. A general approach to derive the constraints for any $Z^{\prime}$ model is discussed in Refs.~\cite{Falkowski:2015krw,Buras:2021btx} which provides the formula used in our analysis. In the limit in which the only quark couplings of the $Z^\prime$ are to $b-s$, LHC searches are not very constraining and assuming a lower limit of $2\,$TeV is not in conflict with ATLAS and CMS searches.

\begin{figure}[t]
	\centering 
	\includegraphics[width=0.45\textwidth]{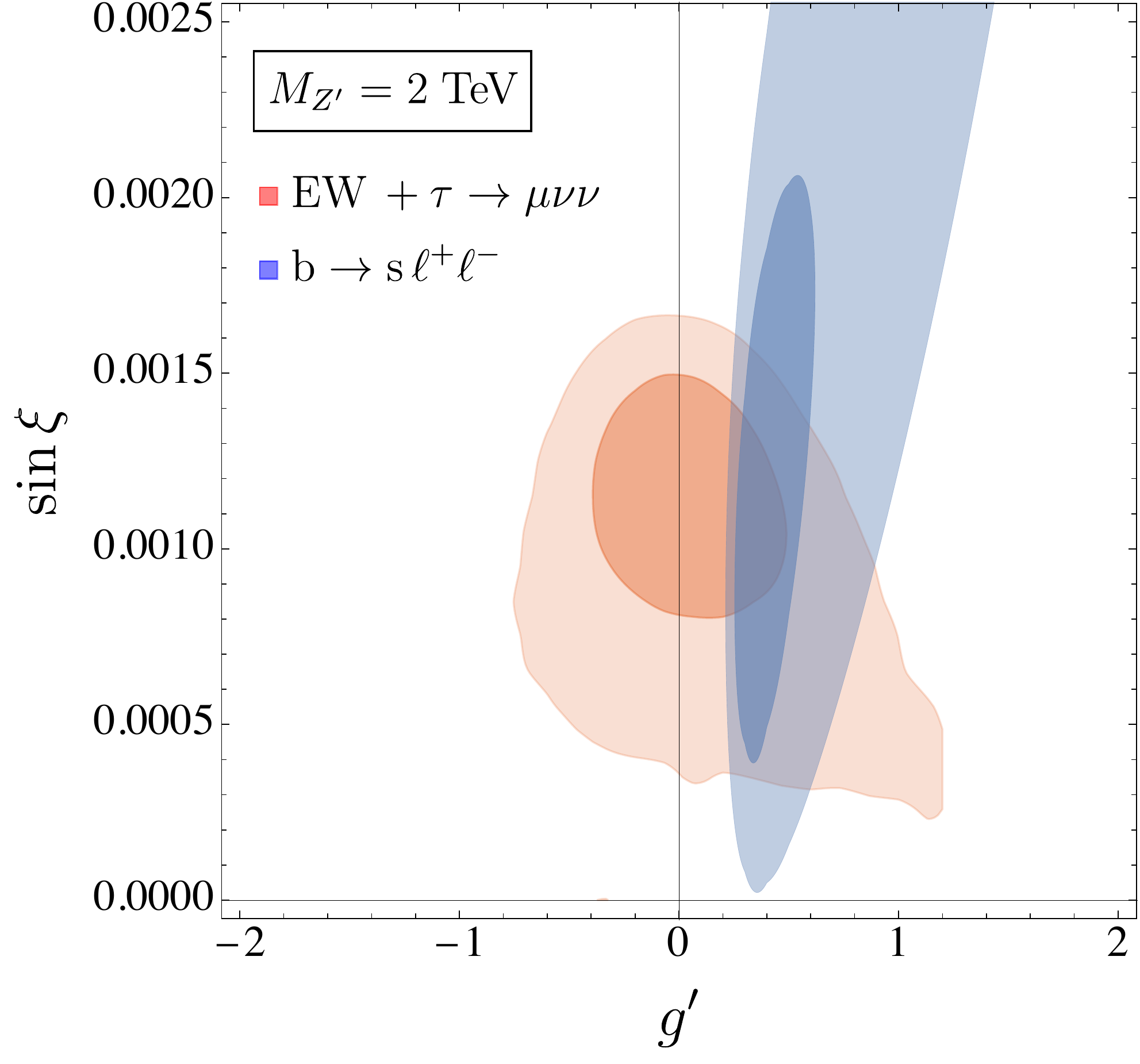}\\
	\includegraphics[width=0.45\textwidth]{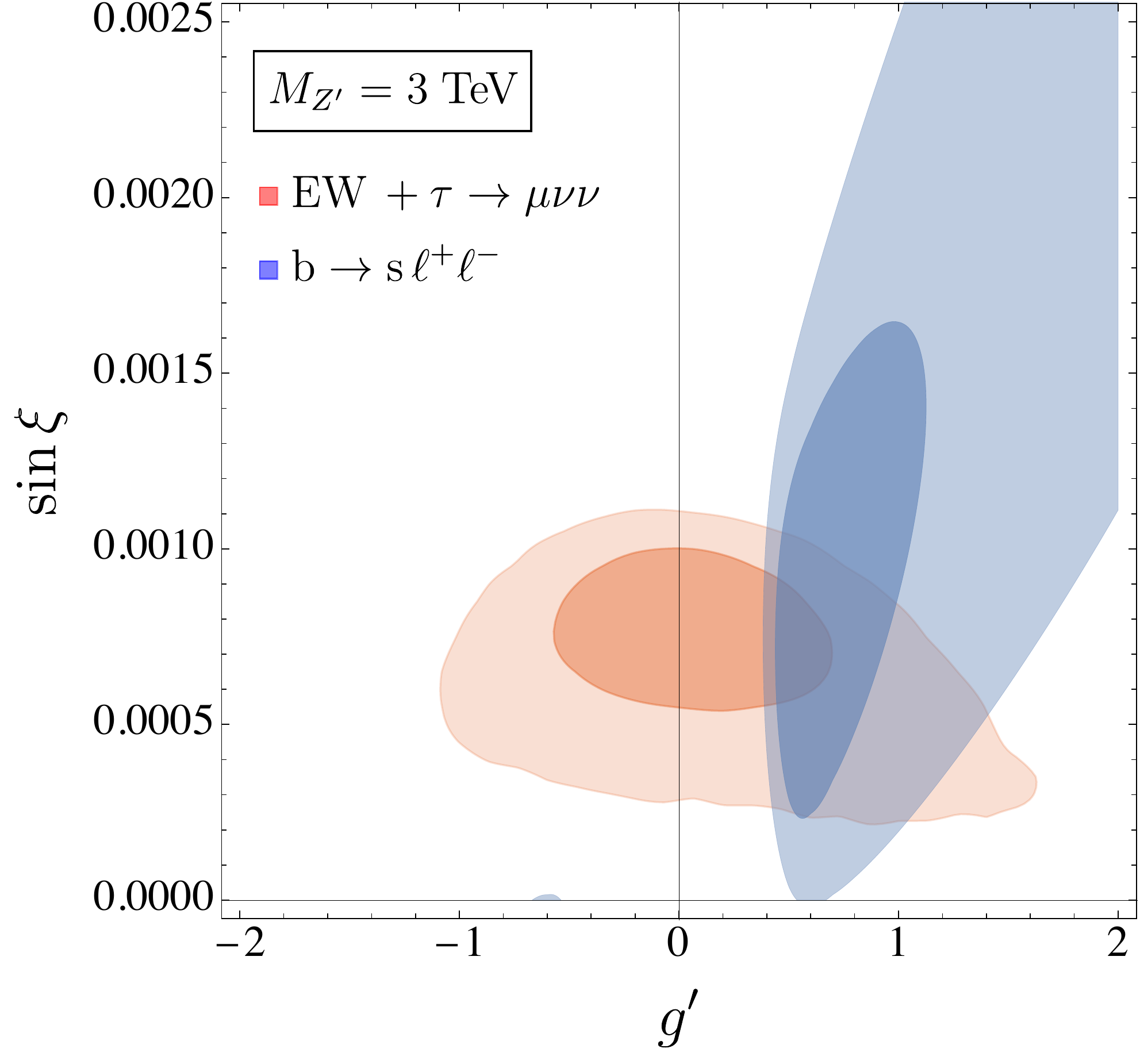}
	\caption{Global fit EW precision observables, neutrino trident production, LEP bounds on 4-lepton contact interactions and $\tau\to\mu\nu\nu$ data (orange) and $b\to s\ell^+\ell^-$ data (blue) in the $g^{\prime}$ - $\sin\xi$ plane for $m_{Z^\prime} = 2\; \rm TeV$ and $m_{Z^\prime} = 3\; \rm TeV$. One can see that both regions overlap nicely and that a non-zero value of the mixing angle is preferred.}%
	\label{fig:2Dfit}%
\end{figure}

\begin{table}[h]
	\begin{tabular}{lccc}
		\toprule 
		 Observable & Scenario 1 & Experiment & Pull  \\
		\hline
		$R_{K^+}^{[1.1,6]}$ & $+0.79\pm 0.01$ & $+0.85\pm 0.04$ & $-1.3$  \\
		$R_{K^0_S}^{[1.1,6]}$ & $+0.79\pm 0.01$ & $+0.66 \pm 0.20$ & $+0.7$  \\
		$R_{K^*0}^{[1.1,6]}$ & $+0.87\pm 0.08$ & $+0.69 \pm 0.12$ & $+1.3$  \\
		$R_{K^{*+}}^{[0.045,6]}$ & $+0.84 \pm 0.04$ & $+0.70 \pm 0.18$ & $+0.8$  \\
		$Q_{5}^{[1.1,6]}$ & $+0.28 \pm 0.02$ & $+0.66 \pm 0.50$ & $-0.8$  \\
		$\langle P_5^\prime \rangle^{[4,6]}$ & $-0.57\pm 0.11$ & $-0.44 \pm 0.12$ & $-0.8$ \\
		$\langle P_5^\prime \rangle^{[6,8]}$ & $-0.79\pm 0.11$ & $-0.58 \pm 0.09$ & $-1.4$  \\
		$10^7 \times {\cal B}_{B_s \to \phi\mu^+\mu^-}^{[4,6]}$ & $+0.78\pm 0.15$ & $+0.62 \pm 0.06$ & $+1.0$  \\
		$10^9\times{\cal B}_{B_s \to \mu^+\mu^-}$ & $+3.08\pm 0.14$ & $+2.85\pm 0.34$ & $+0.6$  \\
		\hline
	\end{tabular} 
	\caption{Predictions for some of the most relevant observables in the $b\to s\ell^+\ell^-$ fit within the scenario of Eq.~\eqref{bsll_sc}. The pulls are given in units of standard deviations. }
	\label{Obs_predictions}
\end{table}
\section{Phenomenology}
\label{Phenomenology}

Let us now study the combined phenomenological consequences of $Z-Z^{\prime}$ mixing in rare semi-leptonic $B$ decays and the global EW fit with the aim of obtaining a combined explanation. For this purpose we will focus on an illustrative simplified scenario with an $SU(2)_L$ singlet $Z^\prime$, such that $Z-Z^\prime$ mixing can account for the discrepancy in the $W$ mass. Furthermore, $b\to s\ell^+\ell^-$ data motivates vectorial couplings to leptons, i.e. $g_{ii}^{\ell L} = g_{ii}^{\ell R} = g_{ii}^{\ell V}$ which also allow for simple configurations without gauge anomalies such as $L_\mu-L_\tau$~\cite{Altmannshofer:2014cfa,Crivellin:2015era} or $B_3-L_2$~\cite{Allanach:2020kss}. In addition $g_{11}^{\ell V}=0$ and $g_{22}^{\ell V}=-g_{33}^{\ell V}=g^\prime$, i.e.~a $L_\mu=-L_\tau$ symmetry~\cite{He:1990pn,Foot:1990mn,He:1991qd}, is motivated by the EW fit since the effect of $Z-Z^\prime$ mixing in $Z\to\nu\nu$ will cancel to leading order. Therefore, larger lepton couplings are possible (see Fig.~\ref{fig:EWfit}) and $\tau\to\mu\nu\nu$ receives the desired constructive contribution via $W-Z^\prime$ box diagrams.\footnote{Note that our analysis would to a good approximation also apply to other scenarios, such as $B_3-L_2$.} In addition to these couplings to leptons, we assume only the presence of left-handed $Z^{\prime}-b-s$ couplings.\footnote{Note that such a scenario could be generated in models with vector-like quarks~\cite{Altmannshofer:2014cfa,Bobeth:2016llm} where absence of $Z^\prime$ couplings to light quarks avoids problems with direct LHC searches as well as larger effects in the total $Z$ width from mixing.}

Importantly, as discussed in the introduction, the current experimental average for the mass of the $W$ boson, shows at least a $3.7~\sigma$ discrepancy with the value predicted from the EW fit within the SM~\cite{deBlas:2022hdk}. This prediction is changed in our model according to \eq{eq:EW:Wmass} such that one accounts for data with a non-zero mixing angle of $|\sin\xi| \simeq 3.5\times10^{-3} \times 1{\rm TeV}/m_{Z^\prime}$. Moving to the complete EW fit (including also LFUV in tau decays, LEP bounds on 4-lepton operators and neutrino trident production) we have $m_{Z^{\prime}},\, g^{\prime}$ and $\sin\xi$ as free parameters. However, since all expressions depend on $g^{\prime 2}/m_{Z^\prime}^2$ despite logarithmic terms we set $m_{Z^\prime}=2\,$TeV. The resulting preferred regions from the EW fit and LFUV in tau decays are shown in Fig.~\ref{fig:2Dfit} for $m_{Z^\prime}=2$ and $3$ TeV. Including $b\to s\ell^+\ell^-$ as well as $B_s-\bar B_s$ mixing, in addition $g^{dL}_{23}$ enters as a free parameter. Marginalizing over $g^{dL}_{23}$ we find the $1\,\sigma$ and $2\,\sigma$ regions shown in blue in Fig.~\ref{fig:2Dfit}. Note that all $2\sigma$ regions nicely overlap, showing that both the EW fit and $b\to s\ell^+\ell^-$ data prefer a non-zero $Z-Z^\prime$ mixing angle such that the $W$ mass can be explained.

%
\section{Conclusions and Outlook}\label{sec:conclusions}

In this article we systematically studied the impact of $Z-Z^\prime$ mixing on the global fit to $b\to s\ell^+\ell^-$ data and EW precision observables. Concerning the former, we observe that a LFU effect is generated while in the latter the mixing leads to modified $Z$ couplings and to an enhancement in the predicted $W$ mass w.r.t.~the SM, which accommodates the {new experimental average} (including the recent one from CDF). Therefore, while in previous analyses in the literature the effect of $Z-Z^\prime$ mixing was usually assumed to be small and was therefore mostly neglected, we stress that both $b\to s\ell^+\ell^-$ data and the EW fit even prefers a small but non-zero value of the order of $10^{-3}$ {for $m_{Z^\prime}\approx 1{\rm TeV}-5 {\rm TeV}$}. Note that this is in agreement with the expectation $\sin \xi\approx g_2 g^\prime m_Z^2/m_{Z^\prime}^2$ for a TeV scale $Z^\prime$ with order one couplings in case $U(1)^\prime$ and EW symmetry breaking are related.

If $b\to s\ell^+\ell^-$ data is in fact explained by a $Z^\prime$ with non-vanishing $Z-Z^\prime$ mixing, one predicts a pattern for the main observables driving the anomaly as shown in Table~\ref{Obs_predictions}. We observe that all tensions with experiment reduce significantly below the 1.5$\,\sigma$ level in the scenario analyzed. Because $b\to s \ell^+\ell^-$ ratios testing LFUV depend naturally (and almost entirely) on ${\cal C}_{9\mu}^{\rm V}$ and thus do not carry information on $\sin\xi$, angular observables are necessary for a distinctive study of $Z^\prime$ models. It will therefore be important to verify with more precise LHCb data together with future Belle~II analysis if this scenario gets reinforced.

Furthermore, forthcoming LHC measurements of the $W$ mass may reinforce the current tension and any improvement in the global EW fit (e.g. in the top mass or in $Z$ decays) would lead to a more precise $W$ mass predictions which could be very precisely measured with future lepton colliders such as FCC-ee~\cite{Abada:2019zxq}, ILC~\cite{Baer:2013cma}, CEPC~\cite{An:2018dwb} or CLIC~\cite{Aicheler:2012bya}. Importantly, if a non-zero $Z-Z^\prime$ mixing is established in the future, {like e.g.~predicted in the model of Ref.~\cite{Crivellin:2015lwa},} this would imply that $SU(2)_L$ and $U(1)^\prime$ are broken by a field charged under both symmetries with important consequences for model building.

\begin{acknowledgements}
We thank Joe Davighi for bringing a missing factor 1/2 in Eq.~(\ref{eq6}) to our attention.
 The work of A.C. and C.A.M. is supported by a Professorship Grant (PP00P2\_176884) of the Swiss National Science Foundation. JM  gratefully acknowledges the financial support by ICREA under the ICREA Academia programme. JM and MA received financial support from Spanish Ministry of Science, Innovation and Universities (project PID2020-112965GB-I00/AEI/ 10.13039/501100011033) and from the Research Grant Agency of the Government of Catalonia (project SGR 1069).	 
\end{acknowledgements}

\section*{Appendix}

\begin{table}[t]
\centering
	\begin{tabular}{l | r } \hline
		Observable & Experimental value \\
		\hline
		$m_W\,[\text{GeV}]$  & $80.379(12)$  \\
		$\Gamma_W\,[\text{GeV}]$  & $2.085(42)$ \\
		${\cal B}(W\to \text{had})$  & $0.6741(27)$ \\
		${\cal B}(W\to \text{lep})$  & $0.1086(9)$ \\
		$\text{sin}^2\theta_{\rm eff,\, e}^{\rm CDF}$ & $0.23248(52)$  \\
		$\text{sin}^2\theta_{\rm eff,\, e}^{\rm D0}$ & $0.23146(47)$  \\
		$\text{sin}^2\theta_{\rm eff,\, \mu}^{\rm CDF}$ & $0.2315(10)$  \\
		$\text{sin}^2\theta_{\rm eff,\, \mu}^{\rm CMS}$ & $0.2287(32)$  \\
		$\text{sin}^2\theta_{\rm eff,\, \mu}^{\rm LHCb}$ & $0.2314(11)$  \\
		$P_{\tau}^{\rm pol}$  &$0.1465(33)$  \\
		$A_{e}$  &$0.1516(21)$ \\
		$A_{\mu}$  &$0.142(15)$ \\
		$A_{\tau}$  &$0.136(15)$ \\		
		$\Gamma_Z\,[\text{GeV}]$  &$2.4952(23)$  \\
		$\sigma_h^{0}\,[\text{nb}]$  &$41.541(37)$ \\
		$R^0_{e}$  &$20.804(50)$ \\
		$R^0_{\mu}$  &$20.785(33)$ \\
		$R^0_{\tau}$  &$20.764(45)$ \\
		$A_{\rm FB}^{0,e}$ &$0.0145(25)$  \\
		$A_{\rm FB}^{0,\mu}$ &$0.0169(13)$  \\
		$A_{\rm FB}^{0,\tau}$ &$0.0188(17)$  \\
		$R_{b}^{0}$  &$0.21629(66)$  \\
		$R_{c}^{0}$  &$0.1721(30)$  \\
		$A_{\rm FB}^{0,b}$  &$0.0992(16)$   \\
		$A_{\rm FB}^{0,c}$  &$0.0707(35)$  \\
		$A_{b}$  &$0.923(20)$  \\
		$A_{c}$  &$0.670(27)$  \\
	\end{tabular}
	\caption{ Electroweak observables~\cite{ALEPH:2005ab,Zyla:2020zbs} used in our fit performed using HEPfit~\cite{deBlas:2019okz} with $m_{Z_0}$, $\alpha$ and $G_F$ as input.}
	\label{tab:EWCC}
\end{table}

We write the interactions of the SM $Z$ with fermions as: 
\begin{align}
	\begin{split}
	\mathcal{L}_{Zff}=&\;\overline{\ell}_j \gamma_\mu \left(\Delta_{ji}^{\ell L} P_L+\Delta_{ji}^{\ell R} P_R\right) \ell_i Z^\mu +\overline{\nu}_j \gamma_\mu \Delta_{ji}^{\nu L} P_L\nu_i Z^\mu\\
	&+\overline{u}_j\,\gamma_\mu \left(\Delta_{ji}^{u L} P_L+\Delta_{ji}^{u R} P_R\right)u_i\,Z^\mu\\
	&+\overline{d}_j\,\gamma_\mu \left(\Delta_{ji}^{d L} P_L+\Delta_{ji}^{d R} P_R\right) d_i\,Z^\mu\,, 
	\end{split}
\end{align}
with $i,j = 1,2,3$ and

	\begin{align}
		\begin{split}
		&\Delta_{ji}^{\ell L(R)} \simeq \sin \xi \, g_{ji}^{\ell L(R)}+\,g_{\rm SM}^{\ell L(R)}\delta_{ji}\,,\\
		&\Delta_{ji}^{\nu L} \simeq \sin \xi \, g_{ji}^{\ell L}+\,g_{\rm SM}^{\nu L}\delta_{ji}\,,\\
		&\Delta_{ji}^{u L} \simeq \sin \xi \, V_{jk} g_{kk'}^qV_{ik'}^*+\,g_{\rm SM}^{u L}\delta_{ji}\,,\\
		&\Delta_{ji}^{u R} \simeq \sin \xi \, g_{ji}^{u}+\,g_{\rm SM}^{u R}\delta_{ji}\,,\\
		&\Delta_{ji}^{d L(R)} \simeq \sin \xi \, g_{ji}^{q (d)}+\,g_{\rm SM}^{d L(R)}\delta_{ji}\,,\\
		\end{split}
		\label{Deltas}
	\end{align}
where $g_{\rm SM}^{i L(R)}$ are the SM couplings given by 
\begin{eqnarray}
		g_{\rm SM}^{\nu L} &=&-\frac{e}{2s_W c_W}\,, \nonumber\\
		g_{\rm SM}^{\ell L} &=&\frac{e}{2s_W c_W}\left(1-2 s_W^2\right)\,,\qquad
		g_{\rm SM}^{\ell R} =-\frac{e \,s_W}{c_W}\,, \nonumber\\
		g_{\rm SM}^{uL}&=&-\frac{e}{s_W c_W}\left(\frac{1}{2}-\frac{2}{3}s_W^2\right)\,,\qquad
		g _{\rm SM}^{uR}=\frac{2}{3}\frac{e\,s_W}{c_W}\,,\,\,\,\,\,\,\,\\
		g _{\rm SM}^{dL}&=&\frac{e}{s_W c_W}\left(\frac{1}{2}-\frac{1}{3}s_W^2\right)\,,\qquad
		g _{\rm SM}^{dR}=-\frac{1}{3}\frac{e \,s_W}{c_W}\,, \nonumber
		\label{eq:SMcouplings} 
\end{eqnarray}

{with $e=g_1 g_2/\sqrt{g_1^2+g_2^2} = g_1 c_W = g_2 s_W$ being} the electric charge.
Moreover, taking into account the $Z-Z^\prime$ mixing in Eq.~(\ref{Deltas}) and the vertex corrections~\cite{Altmannshofer:2014cfa,Haisch:2011up}, we have the following modified $Z$ couplings to leptons
\begin{equation}
\begin{aligned}
\Delta_{ij}^{\ell L} &= g_{\rm SM}^{\ell L}\left( \delta_{ij} + \sin \xi \frac{g_{ij}^{\ell L}}{g_{\rm SM}^{\ell L}} + \sum\limits_k \frac{g_{ik}^{\ell L} g_{kj}^{\ell L}}{(4\pi )^2} {K_F}\left( \frac{m_Z^2}{m_{Z'}^2} \right) \right)\,,
\\
\Delta_{ij}^{\nu L} &= g_{\rm SM}^{\nu L}\left( {{\delta_{ij}} + \sin \xi \frac{{g_{ij}^{\ell L}}}{{g_{{\rm{SM}}}^{\nu L}}} + \sum\limits_k {\frac{{g_{ik}^{\ell L}g_{kj}^{\ell L}}}{{{{(4\pi )}^2}}}} {K_F}\left( {\frac{{m_Z^2}}{{m_{Z'}^2}}} \right)} \right)\,,
\\
\Delta_{ij}^{\ell R} &= g_{\rm SM}^{\ell R}\left( {{\delta_{ij}} + \sin \xi \frac{{g_{ij}^{\ell R}}}{{g_{{\rm{SM}}}^{\ell R}}} + \sum\limits_k {\frac{{g_{ik}^{\ell R} g_{kj}^{\ell R}}}{{{{(4\pi )}^2}}}} {K_F}\left( {\frac{{m_Z^2}}{{m_{Z'}^2}}} \right)} \right)\,,
\end{aligned}
\label{Deltas_with_loops}
\end{equation}
at the $Z$ pole with 
\begin{align}
\begin{split}
{K_F}\left( x \right) =&  - \frac{{2{{(x + 1)}^2}({\rm{L}}{{\rm{i}}_2}( - x) + \ln (x)\ln (x + 1))}}{{{x^2}}} \\
& - \frac{{7x + 4}}{{2x}} + \frac{{(3x + 2)\ln (x)}}{x}.
\end{split}
\end{align}

The effective Hamiltonian~\cite{Grinstein:1987vj,Buchalla:1995vs} in which heavy degrees of freedom have been integrated out is given by:
\begin{equation}
{\cal H}_{\rm eff}=-\frac{4G_F}{\sqrt{2}} V_{tb}V_{ts}^*\sum_i \C{i}  {\cal O}_i
\end{equation}
The relevant operators for this paper are:
\begin{eqnarray}
\begin{aligned}
{\mathcal{O}}_{9\ell} &= \frac{e^2}{16 \pi^2} 
(\bar{s} \gamma_{\mu} P_L b)(\bar{\ell} \gamma^\mu \ell)\,,\\
{\mathcal{O}}_{{9\ell}^\prime} &= \frac{e^2}{16 \pi^2} 
(\bar{s} \gamma_{\mu} P_R b)(\bar{\ell} \gamma^\mu \ell) \,, \\
{\mathcal{O}}_{10\ell} &=\frac{e^2}{16 \pi^2}
(\bar{s}  \gamma_{\mu} P_L b)(  \bar{\ell} \gamma^\mu \gamma_5 \ell)\, ,\\
{\mathcal{O}}_{{10\ell}^\prime} &=\frac{e^2}{16\pi^2}
(\bar{s}  \gamma_{\mu} P_R b)(  \bar{\ell} \gamma^\mu \gamma_5 \ell) \,,
\end{aligned}
\end{eqnarray}
where $P_{L,R}=(1 \mp \gamma_5)/2$. 

The set of observables used in the EW fit are given in Table~\ref{tab:EWCC}.

\bibliography{bibliography}
\end{document}